\documentclass{article}
\RequirePackage{filecontents}
\PassOptionsToPackage{dvipsnames}{xcolor}
\RequirePackage{comgipp}
\RequirePackage{amsfonts}
\RequirePackage{amsmath}
\RequirePackage{authblk}
\usepackage[utf8]{inputenc}
\usepackage[T1]{fontenc}

\newcommand{\theReferenceText}{\fullcite{Scharpf2021}}
\usepackage[
backend=biber,
style=numeric,
firstinits=true,
url=false,
isbn=false,
safeinputenc,%
]{biblatex}
\usepackage{subcaption}
\title{Fast Linking of Mathematical Wikidata Entities in Wikipedia Articles Using Annotation Recommendation}
\author[1]{Philipp Scharpf}
\author[2]{Moritz Schubotz}
\author[3]{Bela Gipp}
\affil[1]{University of Konstanz, Germany (\{first.last\}@uni-konstanz.de)}
\affil[2]{FIZ-Karlsruhe, Germany (\{first.last\}@fiz-karlsruhe.de)}
\affil[3]{University of Wuppertal, Germany ( \{last\}@uni-wuppertal.de)}
\begin{document}
  \maketitle
  \thispagestyle{firststyle}
  \begin{abstract}
Mathematical information retrieval (MathIR) applications such as semantic formula search and question answering systems rely on knowledge-bases that link mathematical expressions to their natural language names.
For database population, mathematical formulae need to be annotated and linked to semantic concepts, which is very time-consuming.
In this paper, we present our approach to structure and speed up this process by supporting annotators with a system that suggests formula names and meanings of mathematical identifiers.
We test our approach annotating 25 articles on \url{en.wikipedia.org}.
We evaluate the quality and time-savings of the annotation recommendations.
Moreover, we watch editor reverts and comments on Wikipedia formula entity links and Wikidata item creation and population to ground the formula semantics.
Our evaluation shows that the AI guidance was able to significantly speed up the annotation process by a factor of 1.4 for formulae and 2.4 for identifiers.
Our contributions were reverted in 12\% of the edited Wikipedia articles and 33\% of the Wikidata items within a test window of one month.
The >>AnnoMathTeX<< annotation recommender system is hosted by Wikimedia at \url{annomathtex.wmflabs.org}.
In the future, our data refinement pipeline is ready to be integrated seamlessly into the Wikipedia user interface.
  \end{abstract}

\section{Introduction}\label{sec:intro}

\begin{figure}
    \includegraphics[width=\columnwidth]{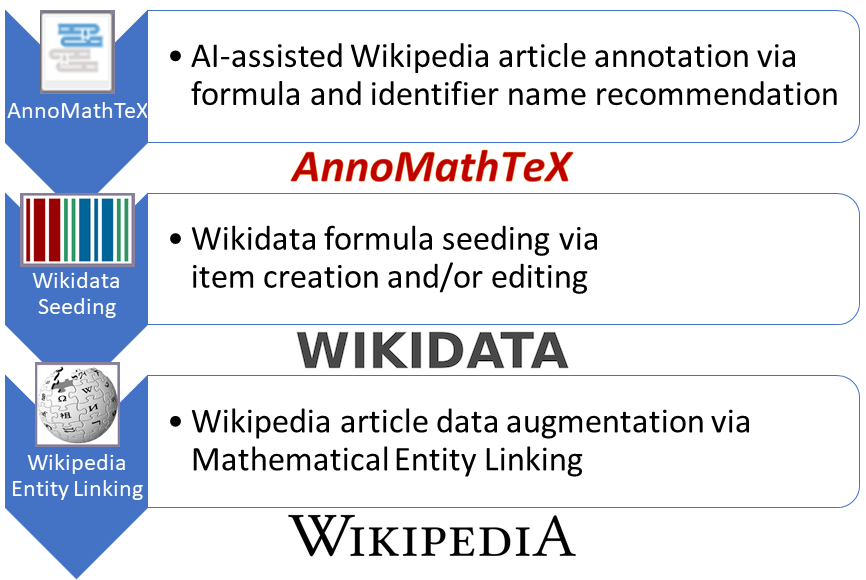}
	\caption{Three-step data refinement pipeline summarizing the contributions of this paper.}
	\label{fig:DataRefinementPipeline}
\end{figure}

In Mathematical Information Retrieval (MathIR), a variety of information systems depend on high-quality data of annotated mathematical formulae to address the human information need.
The human reader is increasingly assisted by systems that enhance document or article readability by providing or linking to additional information.
In the case of semi-structured Wikipedia articles, Wikimedia launched an additional structured database, Wikidata, for a language-independent grounding of concept entities~\cite{DBLP:journals/cacm/VrandecicK14}.
Mathematical Entity Linking (MathEL) is a method to enrich mathematical documents by linking formulae with their constituting entities (identifiers, operators, etc.) to concept representations in knowledge-bases such as Wikipedia or Wikidata~\cite{DBLP:conf/icadl/KristiantoTA16,DBLP:conf/wsdm/KristiantoA17}.
Linking formula concept entities is useful to make this extra information accessible and allow for formula referencing (math citations).
MathIR systems, such as computer algebra systems (CAS), question answering systems (MathQA), recommender systems, plagiarism detection systems (MathPD), or document classification systems, can then exploit the semantic, machine-interpretable data of enriched articles and formulae.
Having motivated the demand or `why' of MathEL, the remaining question is `How can we obtain a large dataset of structured semantic linked formula data?'. Analyzing the statistics of the Wikidata item seeding history, it is evident that the endeavor so far has been an uncoordinated process (various interests) with slow progress (less than 5000 items in 12 years).
In this paper, we report first advances in our project to structure and speed up this dataset creation process.
Our contribution is two-fold: 1) we structure the Wikidata item population by employing Wikipedia article link necessity as motivating selection criterion (application-driven), and 2) we facilitate and accelerate the process by employing a recommender system for formula and identifier annotation and linking (AI-aided).
The system would even allow for unsupervised fully-automatic annotations but we first start with supervised semi-automatic annotations since we consider human quality assessment to be an important control mechanism.
To achieve our research goals, we carry out a three-step pipeline (Figure \ref{fig:DataRefinementPipeline}).
First, we assign formula and identifier names in selected Wikipedia articles using our >>AnnoMathTeX<< system\footnote{A demovideo is available at \url{purl.org/annomathtex}.} that was recently introduced~\cite{DBLP:conf/recsys/ScharpfMSBBG19}.
Second, we create Formula Concept items in the Wikidata knowledge-base.
Third, we integrate our annotations using Entity Linking in Wikipedia articles to our previously created Wikidata items.
To evaluate our data enrichment pipeline, we perform the following research tasks:
\begin{enumerate}
    \item We evaluate the acceptance rate and speedup of the AI assistance.
    \item We evaluate the community acceptance of Wikipedia article formula concept entity links in terms of accepted changes and issue comments.
    \item We evaluate the community acceptance of Wikidata item creation and population in terms of accepted changes and issue comments.
\end{enumerate}

\section{Related Work}\label{sec:related.work}

In this section, we describe the state-of-the-art in the topics that are relevant to our contribution.

\paragraph{Wikification}

Named Entities are typically grounded to a reference object in a knowledge-base.
If Wikipedia articles or Wikidata items are linked, the enrichment is called `Wikification'.
Hachey et al. compare different strategies for Entity Linking to Wikipedia articles that employ candidate identification, disambiguation, coreference, and acronym handling and ranking~\cite{DBLP:journals/ai/HacheyRNHC13}.
Geiss et al. introduce a Named Entity Classifier for Wikidata (`NECKAr') that assigns Wikidata items to the three main NE classes (person, organization, and location), and a Wikidata NE dataset containing over 8 million classified entities~\cite{DBLP:conf/gldv/GeissSG17}.

\paragraph{Mathematical Entity Linking}

A mathematical formula consists of operators, identifiers, and numbers that can be denoted using the Mathematical Markup Language (MathML)\footnote{\url{https://www.w3.org/Math}}. The \textit{LaTeXML} converter\footnote{\url{https://dlmf.nist.gov/LaTeXML}} constructs MathML markup from a LaTeX formula string. Once marked, the identifiers still need to be disambiguated since the same character can have a multitude of different meanings, e.g., $E$ can denote energy, expectation value, etc. There have been efforts to automatically retrieve the semantics of identifiers from the surrounding text~\cite{DBLP:conf/sigir/SchubotzGLCMGYM16}.
A benchmark MathMLben~\cite{DBLP:conf/jcdl/SchubotzGSMCG18} was created containing formulae from Wikipedia, the arXiV and the DLMF, which were augmented by Wikidata markup~\cite{DBLP:conf/sigir/ScharpfSG18}. Greiner-Petter and Schubotz~\cite{DBLP:conf/www/Greiner-PetterS20} examine distributions of mathematical notation on two large corporae from the arXiv\footnote{\url{https://arxiv.org}} and zbMATH\footnote{\url{https://zbmath.org}} repository. They discover `Mathematical Objects of Interest (MOI)', which are potential candidates for MathEL.
Kristianto et al. propose methods to link mathematical expression in scientific documents to Wikipedia articles using their surrounding text~\cite{DBLP:conf/icadl/KristiantoTA16,DBLP:conf/wsdm/KristiantoA17}. Their learning-based approach achieves a precision of 83.40\%, compared with a 6.22  baseline of a traditional MathIR method. A balanced combination of mathematical and textual elements is required for the linking performance to be reliable.
Besides linking to Wikipedia, Schubotz, Scharpf et al.~\cite{DBLP:conf/jcdl/SchubotzGSMCG18,DBLP:conf/sigir/ScharpfSG18} describe linking mathematical formula content to Wikidata, both in MathML and \LaTeX markup. To extend classical citations by mathematical, they call for a `Formula Concept Discovery (FCD) and Formula Concept Recognition (FCR) challenge' to elaborate automated MathEL. Their FCD approach yields a recall of 68\% for retrieving equivalent representations of frequent formulae, and 72\% for extracting the formula name from the surrounding text on the NTCIR arXiv dataset~\cite{DBLP:conf/ntcir/AizawaKOS14}.
Entity linking has a wide range of IR and NLP applications, such as semantic search and question answering, text enrichment, relationship extraction, entity summarization, etc.~\cite{DBLP:conf/amw/Rosales-MendezP18}.
Mathematical Entity Linking - being less popular than its natural language correspondent - has so far been employed in mathematical question answering systems, such as `MathQA' using structured Wikidata items~\cite{DBLP:journals/corr/abs-1907-01642} and proposed for semi-structured question posts from Math Stack Exchange (MSE) at the CLEF ARQMath Lab~\cite{DBLP:conf/clef/ScharpfSGOTG20}. Moreover, it is expected that MathEL will enhance mathematical subject classification~\cite{DBLP:conf/jcdl/ScharpfSYHMG20,DBLP:conf/mkm/SchubotzSTKBG20}.

\paragraph{Document Annotation Recommendation}

In the field of document analysis, annotation means adding semantic information, mainly by linking and disambiguating entity references~\cite{DBLP:journals/software/FerraginaS12}. Since disambiguation requires understanding context, human inspection is needed in most cases. To facilitate and speedup the process, annotation recommender systems can be used~\cite{DBLP:conf/recsys/ScharpfMSBBG19}. Previous research has been focused on tag recommendation or suggestion. Musto et al. present a `Social Tag Recommender System (STaR)' for social documents~\cite{DBLP:conf/pkdd/MustoNGLS09}. While their system exploits classical features for text similarity (tf–idf), more recent approaches~\cite{DBLP:conf/icai3/YangNL20} use neurally learned representations (fastText). Additionally, they provide an online user interface for interactive customization of recommendation thresholds. Kowald et al. present a framework (`TagRec')~\cite{DBLP:conf/um/KowaldKL17} that can flexibly employ 10 different metrics and 12 recommendation algorithms trained on 11 datasets (various data types, such as texts, images, music, movies, etc.).

\paragraph{Formula Concept Discovery and Recognition}

Scharpf et al.~\cite{DBLP:conf/sigir/ScharpfSG18,DBLP:conf/sigir/ScharpfSCG19} recently defined a `Formula Concept' as a labeled collection of mathematical formulae that are equivalent but have different representations through notation, e.g., the use of different identifier symbols or commutations. Different notations can make human and machine understanding of mathematical formulae non-trivial. Consider for example the formula $E=mc^2$. It can be regarded as being one representation of a Formula Concept labeled `mass-energy equivalence'. A different representation of this same concept could be $\mu=\epsilon/c^2$.
The challenge of Formula Concept retrieval~\cite{DBLP:conf/sigir/ScharpfSCG19} (a method for MathEL) can roughly be split into the discovery (defining concepts by exploring some instances) of Formula Concepts and their recognition (matching new instances to prior defined concepts represented by name\footnote{Possibly grounded by an item in a semantic knowledge-base, such as Wikidata.}).
A Wikidata Entity Linking markup to for \LaTeX~ and MathML was introduced and discussed in~\cite{DBLP:conf/jcdl/SchubotzGSMCG18} and~\cite{DBLP:conf/sigir/ScharpfSG18}. The proposed markup should be used by authors of documents in the STEM disciplines to semantically annotate mathematical content in documents. For example,\\
\verb|$\w{Q210546}{E=mc^2}$|\\
in \LaTeX~ markup is intended to include a link to the Wikidata item QID\footnote{A unique ID assigned to each Wikidata item.} for the formula $E=mc^2$. This way, the formula string can be assigned to a Formula Concept name - here `mass-energy equivalence'. Furthermore, using the Wikidata annotation, also the constituting identifiers $E$, $m$, and $c$ can be linked to their natural language meanings - here `energy', `mass', and `speed of light (in vacuum)'. The annotation process will gradually result in the creation of a dataset of humanly annotated Formula Concepts and identifiers, which can be used for a variety of IR, AI, and ML applications, as described in the introduction.

In this paper, we start this endeavor in the three-step pipeline, as shown in Figure \ref{fig:DataRefinementPipeline}.

\section{Methods}\label{sec:methods}

In this section, we explain the methods of our Entity Linking pipeline in detail.

\subsection{Wikidata Item Seeding and Semantic Annotation}\label{subsec:wikid.seedanno}

Before annotating and linking formulae in a selection of Wikipedia articles, we need to ground them in the structured knowledge-base Wikidata to allow for Entity Linking in Wikipedia (Section \ref{subsec:wikip.entitylink}). Since Wikidata is open and collaborative, there is already a community creating and populating mathematical formula items. The advantage over most other current Wikidata seeding policies is that our Wikipedia article annotation-driven strategy is structured and application-oriented. In this project, we continue the mathematical question answering (MathQA)~\cite{DBLP:journals/corr/abs-1907-01642} driven seeding. We start by explaining the data model of mathematical Wikidata items.

\begin{figure}
	\includegraphics[width=\columnwidth]{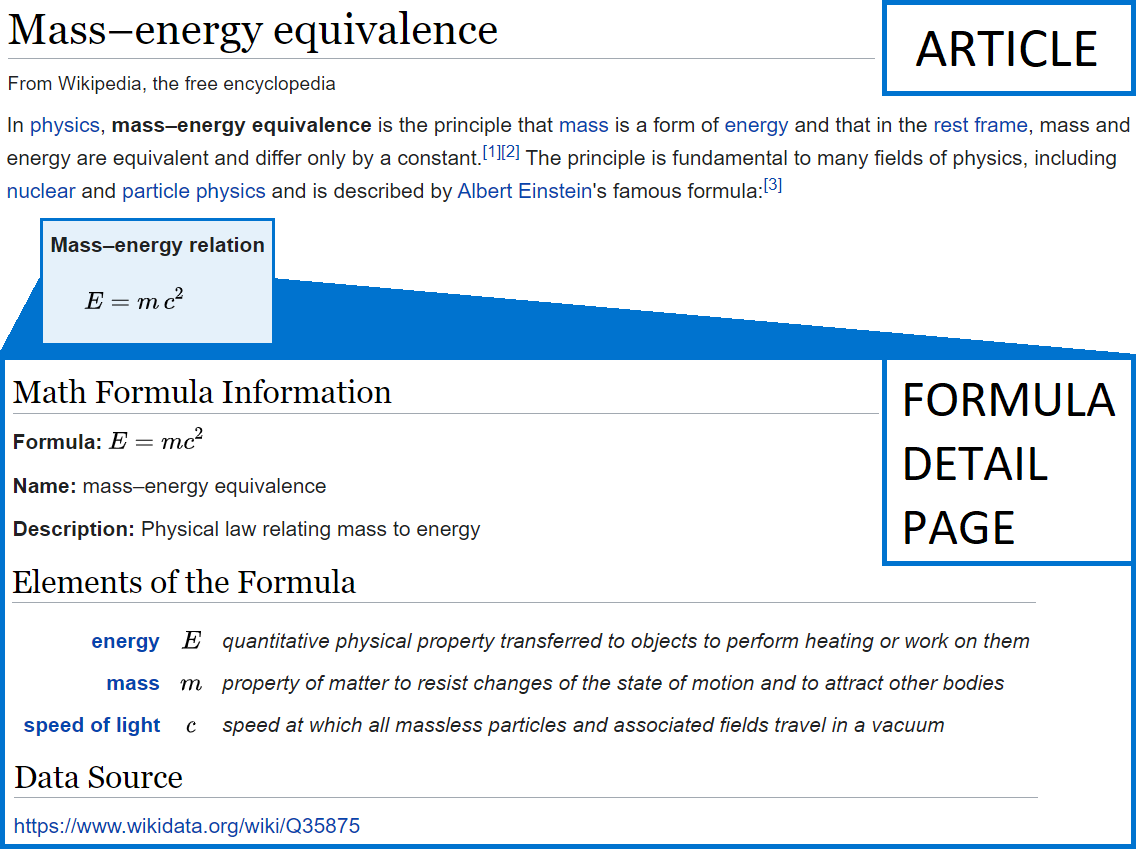}
	\caption{The start of the Wikipedia page on ``mass-energy equivalence'' (above) and the detail page for the linked Formula Concept `mass-energy equivalence' (below). Elements of the formula are retrieved from the `has part' property of the Wikidata item.}
	\label{fig:Emc2}
\end{figure}

For mathematical items, such as `mass-energy equivalence' (Q35875), the \textit{defining formula} property (P2534) is used to store the \LaTeX~ string representation of the formula (here `$E=mc^2$').
Furthermore, the \textit{has part} property (P527) links mathematical Formula Concepts to their constituting identifiers (e.g., \textit{energy}, \textit{mass} and \textit{speed of light} for Q35875).

\subsection{Wikipedia Article Formula Entity Linking}\label{subsec:wikip.entitylink}

If popular Formula Concepts are discovered and seeded into Wikidata with `defining formula' and `has part' properties, they can be employed for various IR applications. In this project, we demonstrate and evaluate Entity Linking in mathematical Wikipedia articles\footnote{A demovideo is available at \url{purl.org/mathwikilink}.}. Figure \ref{fig:Emc2} (above) shows the corresponding Wikipedia article for our `Mass-energy equivalence' example. The central Formula Concept, `mass-energy relation' appears in a blue box, which is linked to the corresponding Wikidata item.
Clicking on the formula block, the reader is linked to a `special page'\footnote{The functionality was introduced in \cite{DBLP:conf/jcdl/SchubotzGMTG20}.} (Figure \ref{fig:Emc2} below) that shows further information about the Formula Concept, retrieved from the corresponding Wikidata item (Q35875), which is declared as `data source' at the bottom. In a section `Elements of the Formula' the meaning (name and description) of the constituting formula identifiers is displayed, retrieved from the item's `has part' property.
To enable Wikimedia Entity Linking, QID attributes must be inserted into <math> tags (for both boxed block-level or inline text formulae)\footnote{In the <math> tag, a distinction is made via a displaystyle attribute.}, following~\cite{DBLP:conf/jcdl/SchubotzGMTG20}. In our example of the first formula in the article `Mass-energy equivalence', this could look like
\begin{verbatim}
<math display="block" qid=Q35875>E=m\,c^2</math>
\end{verbatim}
in the Wikitext source code.
The Wikitext can be edited either manually by users or automated by created bots or pipelines authenticated by user credentials. An example of a popular tool is the python library Pywikibot\footnote{\url{https://www.mediawiki.org/wiki/Manual:Pywikibot}}.

\section{Implementation}\label{sec:implementation}

In the following, we describe the implementation and workflow of our >>AnnoMathTeX<< system that facilitates and accelerates the annotation of mathematical formulae and identifiers in STEM documents by recommending their annotation name candidates (and Wikidata QID if available).

\paragraph{AnnoMathTeX Workflow}

\begin{figure}
    \includegraphics[width=\columnwidth]{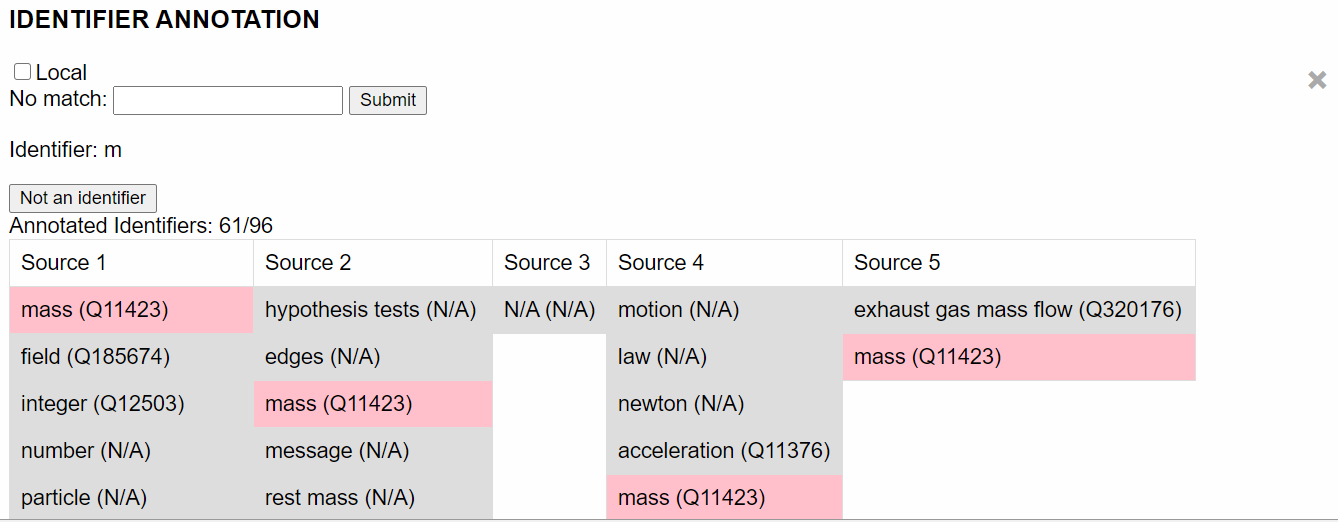}
	\caption{Popup table containing recommendations for the annotation of the identifier `$m$', provided from different sources (cut off after fifth ranked).}
	\label{fig:IdentifierPopup}
\end{figure}

\begin{figure}
	\begin{center}
	\includegraphics[width=\columnwidth]{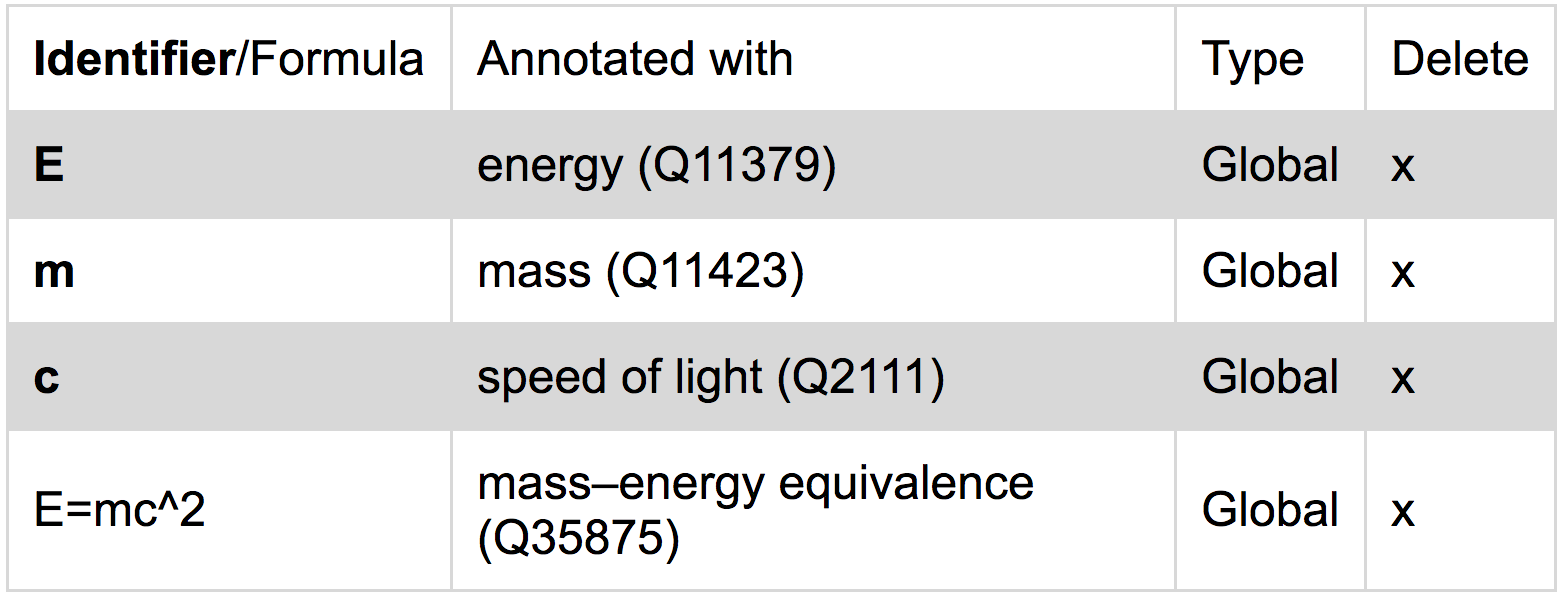}
	\end{center}
	\caption{Content of annotation table after the user has annotated all the identifiers in the formula "$E=mc^2$", as well as the formula itself. Identifiers are written in bold font.}
	\label{fig:AnnoTabEmc2}
\end{figure}

On the start screen, the user can select articles to continue saved annotation sessions or load new articles from Wikipedia. Also, YouTube and GitHub icons link to a tutorial and the project repository, respectively. The user can either download and run the system locally or visit the web version hosted by Wikimedia\footnote{\url{https://annomathtex.wmflabs.org}}.
Besides Wikipedia articles in Wikitext, the system can also parse mathematical documents in LaTeX format. Once the document has been parsed and rendered, the user can click on formulae (delimited by \verb|$| signs) and their constituting identifiers to open the respective annotation recommendation popups. After annotation, the formula delimiters or identifier symbols turn green to visualize the progress, which is reversible.
Figure \ref{fig:IdentifierPopup} shows an example annotation popup for the identifier (`m'). The annotation of formulae can be done analogously.
In both cases, name and Wikidata QID recommendations can be selected from different sources (see next paragraph), which are randomized in their order with anonymized names to avoid bias in the evaluation of their relative performance. In case the identifier or formula was parsed incorrectly, the user can click on the 'Not an identifier/formula' buttons. In case no suitable annotation recommendations are provided, the user can `manually' type in a name. If successful, the annotations are added to an annotations table at the top of the page (see Figure \ref{fig:AnnoTabEmc2}).
All document, annotation, and evaluation files are saved to a separate data repository\footnote{\url{https://github.com/ag-gipp/dataAnnoMathTex}}.

\paragraph{Sources for Recommendations}

The annotation recommendations for identifiers are provided from the following sources:

\begin{itemize}
    \item \textbf{arXiv} - list containing candidate names for all lower- and upper-case Latin and Greek letter identifier symbols appearing in the NTCIR arXiv corpus\footnote{http://ntcir-math.nii.ac.jp/data} - the candidates were extracted from the surrounding text of 60 M formulae and ranked by the frequency of their occurrence;
    \item \textbf{Wikipedia} - occurence frequency ranked list of candidates\footnote{\url{https://en.wikipedia.org/wiki/User:Physikerwelt}} extracted from definitions in mathematical English articles;
    \item \textbf{Wikidata} - dumped list of candidate symbols from formulae with `quantity symbol (string)' property (P416) retrieved via a SPARQL query\footnote{\url{https://query.wikidata.org}};
    \item \textbf{Word window} - list of $\pm 5$ words around the specific formula to be annotated;
    \item \textbf{User input} - saved names that were previously typed in by a user when no matching recommendation was available.
\end{itemize}

The annotation recommendations for formulae are provided from the following sources:

\begin{itemize}
    \item \textbf{Wikidata fuzzy} - string matching with a list of formulae retrieved from Wikidata items with `defining formula' property (P2534);
    \item \textbf{Wikidata parts} - identifier semantics overlap with Wikidata formulae, provided the user has annotated at least one of the identifiers in the formula already;
    \item Formula Concept memory - past formula annotations are stored in the data repository, such that alternative string representations (e.g., \verb|E=mc^2|, \verb|m=E/c^2|, etc.) are collected for identical names - recommendations are provided if name and QID match;
    \item \textbf{Word window} - analogous to word window for identifiers;
    \item \textbf{User input} - analogous to user input for identifiers.
\end{itemize}

\paragraph{Global vs. Local Annotation}

By default, the annotation mode is set to \textit{global}. This means that if the user annotates, e.g., the identifier \textit{E} with \textit{energy}, all other occurrences of the identifier \textit{E} in the document automatically receive this annotation. This way, a significant amount of time is saved. If the usage of the identifier symbols is not consistent within the document (this should not be the case for Wikipedia articles), there is also an option for \textit{local} annotation.

\section{Evaluation}\label{sec:eval}

In this section, we describe evaluation tasks, metrics, and results for our three-step pipeline (article annotation, Wikipedia linking, Wikidata seeding).

\subsection{Dataset Selection}\label{data.sel}

We evaluate the success of the individual steps on a test selection, for which we elaborated several criteria.

\paragraph{Selection Criteria and Sources}

Accounting for the information need of the mathematical community (pupils, students, teachers, researchers), we propose the following selection criteria for Wikipedia articles describing basic physics notions, concepts or equations:

\begin{enumerate}
    \item Popularity: Highest pageview statistics pages\footnote{\url{https://en.wikipedia.org/wiki/Wikipedia:WikiProject_Physics/Popular_pages?oldid=987202322}}\textsuperscript{,}\footnote{\url{https://pageviews.toolforge.org}},
    \item Concepts: Outline\footnote{\url{https://en.wikipedia.org/wiki/Outline_of_physics?oldid=987838936}} of and concepts\footnote{\url{https://en.wikipedia.org/wiki/Category:Concepts_in_physics?oldid=951714261}} in physics pages,
    \item Equations: List of physics equations pages\footnote{\url{https://en.wikipedia.org/wiki/Lists_of_physics_equations?oldid=948648427}}, and
    \item Education: Wikiversity\footnote{\url{https://en.wikiversity.org/wiki/Fundamental_Physics/Formulas?oldid=2188350}} and curicula\footnote{\url{https://en.wikipedia.org/wiki/List_of_physics_concepts_in_primary_and_secondary_education_curricula?oldid=984574720}} pages.
\end{enumerate}

The page snapshots
were taken on November, 18th 2020. The time interval for the pageview statistics (daily averages in Table \ref{tab:SelectedArticles}) is October 2020. From our 7 sources, we selected the 25 most popular (criterion 1) and relevant (criteria 2-4) pages from the physics subfield of `classical mechanics of motion' (linear and rotational)\footnote{\url{https://en.wikipedia.org/wiki/List_of_equations_in_classical_mechanics?oldid=1000494345}}. We apply as filter condition that the selected articles need to describe a physics concept (no person, method or experiment) with at least one block-level formula (physics equation, no chemistry formula).

\paragraph{Article Assessment}

Table \ref{tab:SelectedArticles} shows a list of our 25 selected articles. For each article, we collected its pageview average (daily), importance label (community), and math elements. The latter number was retrieved by searching for all \verb|<math>| tags (formula or identifier environments) in the Wikitext source code. The mean of pageviews is 1747, indicating that many users could potentially profit from our article enhancement via Entity Linking. The mean of math elements number is 72 (total 1797), indicating that the selected articles contain a significant amount of mathematical content.
The articles were downloaded from \url{https://en.wikipedia.org/wiki} on November, 23rd 2020 in >>AnnoMathTeX<< and opened for annotation the following days.

\begin{table}
\caption{Properties (daily pageviews, importance label, math elements) of our 25 selected articles from the physics subfield of `classical mechanics of motion' (linear and rotational), containing a total of 1797 formulae.}
\resizebox{\columnwidth}{!}{
\begin{tabular}{|l|l|l|l|}
\hline
\textbf{Article}                                                     & \textbf{Pageviews} & \textbf{Wikipedia Importance} & \textbf{Math Elements} \\ \hline
Acceleration                  & 1617      & Top        & 47       \\ \hline
Angular\_acceleration         & 452       & N/A          & 37       \\ \hline
Angular\_frequency            & 1081      & High       & 13       \\ \hline
Angular\_momentum             & 1942      & Top        & 189      \\ \hline
Angular\_velocity             & 1441      & Top        & 133      \\ \hline
Center\_of\_mass              & 803       & N/A          & 37       \\ \hline
Centrifugal\_force            & 1158      & High       & 22       \\ \hline
Centripetal\_force            & 1216      & High       & 101      \\ \hline
Circular\_motion              & 1411      & High       & 73       \\ \hline
Coriolis\_force               & 1544      & High       & 48       \\ \hline
Equations\_of\_motion         & 1059      & Mid        & 64       \\ \hline
Force                         & 2161      & Top        & 112      \\ \hline
Frequency                     & 1999      & High       & 15       \\ \hline
Harmonic\_oscillator          & 1101      & Top        & 165      \\ \hline
Jerk\_(physics)               & 625       & N/A          & 30       \\ \hline
Mass                          & 2011      & Top        & 31       \\ \hline
Moment\_of\_inertia           & 2515      & High       & 358      \\ \hline
Momentum                      & 1953      & Top        & 79       \\ \hline
Motion                        & 1138      & Top        & 4        \\ \hline
Newton\%27s\_laws\_of\_motion & 8744      & Top        & 13       \\ \hline
Rotation                      & 441       & N/A          & 42       \\ \hline
Speed                         & 1086      & Top        & 16       \\ \hline
Torque                        & 2454      & High       & 61       \\ \hline
Velocity                      & 1587      & Top        & 33       \\ \hline
Work\_(physics)               & 2130      & High       & 74       \\ \hline
\end{tabular}}
\label{tab:SelectedArticles}
\end{table}

\subsection{Annotation Guidelines and Issues}\label{subsec:ann.guid.iss}

For the annotation, we developed the following rules or guidelines: 1) we annotate identifiers first, such that the formula name recommendation retrieval from Wikidata via the `has part' properties is enabled; 2) we ignore derivative characters, non-relations, tables, derivations and all indices (superscript or subscript); 3) locally different meanings of the same identifier within an article should be avoided (appeal to editors); 4) proper names (e.g., `Planck constant') must be capitalized according to the conventions from `Content dictionary description' (DRMF)~\cite{cohl2017content}. For the full list, see \url{https://github.com/ag-gipp/AnnoMathTeX/guidelines}.

During the annotation process, we discovered the following issues: 1) it is not possible to parse equations with no spaces between identifiers, e.g., in the right-hand side of the \LaTeX~string `\verb|L = rmv|'; 2) there are different common practices to denote vectors in \LaTeX, e.g., \verb|\vec| vs. \verb|\mathbf|; 3) sometimes two names are both commonly used to denote the same Formula Concept, e.g., `M-sigma relation' (Q3424023) and `Faber–Jackson relation' (Q1390162).

\subsection{Recommendations Quality}\label{subsec:rec.qual}

\paragraph{Source Performance Comparison}

Table \ref{tab:SelectionRanking} shows a comparison of the performance of the different annotation recommendation sources (see Section \ref{sec:implementation}). In addition to the ranking of the accepted recommendations (the position at which they appeared in the popup table), the Cumulative Gain (CG) and Discounted Cumulative Gain (DCG) per source are displayed.
The DC and DCG performance measures are calculated according to~\cite{DBLP:journals/tois/JarvelinK02} as
\begin{align*}
    \mathrm{CG_{p}} = \sum_{i=1}^{p} rel_{i}, \quad
    \mathrm{DCG_{p}} = \sum_{i=1}^{p} \frac{rel_{i}}{\log_{2}(i+1)},
\end{align*}
where $rel_i$ is the relevance (here accepted recommendations) at position i and p is the ranking scale cutoff (here position 10).
In contrast to CG, which is simply the total sum of accepted recommendations per source, DCG takes into account the position at which the accepted recommendation appeared. It penalizes low-ranked recommendations by assigning logarithmically decreasing gain.
\begin{table}[ht]
\caption{Source performance comparison for identifiers (above) and formulae (below). The number of times a source was able to provide a name that was accepted by the annotator, and its position in the ranking.}
\resizebox{\columnwidth}{!}{
\begin{tabular}{|l|c|c|c|c|c|c|c|c|c|c|c|c|}
\cline{1-1}
\cline{4-13}
\multicolumn{1}{|l|}{\textbf{Identifiers}} & \multicolumn{1}{}{} & \multicolumn{1}{}{} & \multicolumn{10}{|c|}{\textbf{Position}} \\ \hline
\textbf{Source} & \textbf{CG} & \textbf{DCG} & \textbf{1} & \textbf{2} & \textbf{3} & \textbf{4} & \textbf{5} & \textbf{6} & \textbf{7} & \textbf{8} & \textbf{9} & \textbf{10} \\ \hline \hline
arXiv           & 146 & 111             & 79         & 18          & 20          & 3          & 21          & 2          & 0          & 3          & 0          & 0           \\ \hline
Wikipedia       & 169 & 100             & 45         & 16          & 45          & 15          & 3          & 3          & 18          & 5          & 18          & 1           \\ \hline
Wikidata     & 141 & 85             & 23          & 55          & 4          & 53          & 6          & 0          & 0          & 0          & 0          & 0           \\ \hline
Word window     & 136 & 67             & 14          & 18          & 25          & 20          & 17          & 10          & 7          & 12          & 5          & 8           \\ \hline
\end{tabular}}
\vspace{0.25cm} \newline
\resizebox{\columnwidth}{!}{
\begin{tabular}{|l|c|c|c|c|c|c|c|c|c|c|c|c|}
\cline{1-1}
\cline{4-13}
\multicolumn{1}{|l|}{\textbf{Formulae}} & \multicolumn{1}{}{} & \multicolumn{1}{}{} & \multicolumn{10}{|c|}{\textbf{Position}} \\ \hline
\textbf{Source} & \textbf{CG} & \textbf{DCG} & \textbf{1} & \textbf{2} & \textbf{3} & \textbf{4} & \textbf{5} & \textbf{6} & \textbf{7} & \textbf{8} & \textbf{9} & \textbf{10} \\ \hline \hline
\multicolumn{1}{|l|}{Wikidata fuzzy}      & 18 & 11              & 9          & 0          & 3          & 1          & 0          & 0          & 0          & 0          & 0          & 0           \\ \hline
Wikidata parts      & 11 & 6              & 4          & 3          & 0          & 0          & 0          & 1          & 0          & 0          & 0          & 0           \\ \hline
FC memory            & 66 & 45              & 25          & 11          & 12          & 7          & 2          & 4          & 2          & 2          & 1          & 0           \\ \hline
Word window     & 106 & 67              & 26          & 23          & 12          & 16          & 7          & 7          & 4          & 5          & 3          & 1           \\ \hline
\end{tabular}}
\label{tab:SelectionRanking}
\end{table}
While for identifiers Wikipedia outperformed all other sources in total (CG), the arXiv scored higher considering ranking (DCG).
For formulae, the word window performed best in both CG and DCG measure.
Interestingly, our own Formula Concept (FC) memory, strongly outperformed both Wikidata variants. This indicates a significant global reuse of FCs in our article selection because the content of the articles is semantically closely related. Moreover, this shows that there is this urgent need to seed these FCs into Wikidata (see Section \ref{subsec:form.conc.seed}).
For the majority of the sources (both for identifiers and formulae), the largest amount of recommendations were accepted in the first position. This means that unsupervised semantification (automatically selecting the first-ranked) could potentially be considered.

\paragraph{Wikidata QID Retrieval}

For 80\% of the annotated identifiers, there was a QID available. For the formulae, after the Wikidata seeding (see Section \ref{subsec:form.conc.seed}) 60\% could be attributed to a QID.
For 15 formulae, the disambiguation did not work. For example, 'work' was attributed to Q6958747 (labour) instead of Q42213 (energy transfer). We corrected and inserted them together with the QIDs from our seeding list in the annotation tables.

\paragraph{Average Annotation Time (Recommendations vs. Manual)}

Table \ref{tab:AvgAnnotationTime} shows the average annotation time for identifiers and formulae, respectively, comparing recommendation selection with manual annotation.
The recommendation selection time is measured from the point when the user encounters the annotation recommendations (popup opening) until selection. The manual annotation time is the period in which the user is manually typing in the annotation until the `submit' button is clicked.
\begin{table}
\caption{The average annotation time for identifiers (above) and formulae (below) using recommendation selections vs. manual insertions.}
\begin{center}
\begin{tabular}{l|l|}
\cline{2-2}
\textbf{Identifiers} & Time (seconds)     \\ \hline
\multicolumn{1}{|l|}{Recommendation} & 2.6 \\ \hline
\multicolumn{1}{|l|}{Manual}    & 6.3 \\ \hline
\end{tabular}
\begin{tabular}{l|l|}
\cline{2-2}
\textbf{Formulae} & Time (seconds)     \\ \hline
\multicolumn{1}{|l|}{Recommendation} & 2.8 \\ \hline
\multicolumn{1}{|l|}{Manual}    & 4.0 \\ \hline
\end{tabular}
\end{center}
\label{tab:AvgAnnotationTime}
\end{table}
The results demonstrate that the recommendations lead to significant time savings (factor 2.4 for identifiers and 1.4 for formulae), which accumulate when annotating large corporae.
The manual annotation time depends on the identifier or formula name length and annotator typing speed. For a more slowly typing annotator and long names, the savings are even larger.

\paragraph{Local vs. Global Annotation}

One would expect to need much more identifier than formula annotations, as formulae usually contain multiple identifiers. However, the reuse of identifier annotations per document or even per corpus (article collection) is a significant advantage for the >>AnnoMathTeX<< system to save time through global annotations. If the author's use of identifier symbols is consistent within a document, only one annotation is necessary for all formulae in which the identifier occurs. The time savings are especially large for large documents with many identifier recurrences.

\subsection{Formula Concept Seeding}\label{subsec:form.conc.seed}

Table \ref{tab:SeedingList} shows example lines from the initial seeding list that was generated from the Formula Concept memory and the annotation evaluation files from the >>AnnoMathTeX<< system. The Formula Concept name corresponds to the Wikidata item name. Seed contribution options are item (i), formula (f), parts (p) or a combination thereof. Identifier seeding property options are `has part' (hp) or `calculated from' (cf).
The number of different Formula Concept representation variations (No. FC vars.) is recorded.
After the community feedback from Wikipedia, a second seeding list was necessary to account for the need to seed specific formulae as more fine granular concepts (see Section \ref{subsec:feedb.wikip} and \ref{subsec:cur.gold}).

\begin{table}
\caption{Initial seeding list for Wikidata items (Name, QID) with seeding contribution, number of Formula Concept variations, and identifier property used. Only the cases, where our contributions were needed are shown. For the full list of 66 formulae see the repository.}
\resizebox{\columnwidth}{!}{
\begin{tabular}{|l|l|l|l|l|}
\hline
\textbf{Name}                               & \textbf{QID}        & \textbf{Contrib.} & \textbf{FC vars.} & \textbf{Prop.} \\ \hline
center of mass                     & Q2945123   & p        & 2           & hp    \\ \hline
centripetal acceleration           & Q2248131   & f/p      & 1           & hp    \\ \hline
centripetal force                  & Q172881    & f/p      & 2           & hp    \\ \hline
circumference                      & Q843905    & p        & 1           & hp    \\ \hline
conservation of energy             & Q11382     & f/p      & 2           & hp    \\ \hline
conservation of momentum           & Q2305665   & f/p      & 2           & hp    \\ \hline
damping                            & Q1127660   & f/p      & 1           & hp    \\ \hline
Dirac equation                     & Q272621    & p        & 1           & hp    \\ \hline
Dirac equation in curved spacetime & Q16853908  & p        & 1           & hp    \\ \hline
elastic energy                     & Q891408    & p        & 1           & hp    \\ \hline
electromagnetic force              & Q849919    & f/p      & 2           & hp    \\ \hline
electrostatic force                & Q103438301 & i/f/p    & 1           & hp    \\ \hline
energy-momentum relation           & Q103439852 & i/f/p    & 2           & hp    \\ \hline
escape velocity                    & Q166530    & i/f/p    & 1           & hp    \\ \hline
Euler-Lagrange equation            & Q875744    & p        & 2           & hp    \\ \hline
four-momentum                      & Q1068463   & p        & 1           & hp    \\ \hline
four-velocity                      & Q1322540   & f/p      & 1           & hp    \\ \hline
free fall                          & Q140028    & p        & 1           & hp    \\ \hline
friction                           & Q82580     & f/p      & 1           & hp    \\ \hline
Galilean transformation            & Q219207    & p        & 3           & hp    \\ \hline
gravitational acceleration         & Q30006     & f/p      & 2           & hp    \\ \hline
gravitational force                & Q11412     & f/p      & 2           & hp    \\ \hline
gravitational potential            & Q1544012   & f/p      & 4           & hp    \\ \hline
Hamiltonian operator               & Q660488    & f        & 2           & hp    \\ \hline
Hooke's law                        & Q170282    & p        & 3           & hp    \\ \hline
jerk                               & Q497332    & f/p      & 1           & hp    \\ \hline
Lagrangian operator                & Q103687426 & i/f/p    & 2           & hp    \\ \hline
Lorentz factor                     & Q599404    & f/p      & 5           & hp    \\ \hline
Lorentz force                      & Q172137    & p        & 1           & hp    \\ \hline
Lorentz transformation             & Q217255    & f/p      & 2           & hp    \\ \hline
Newton's third law of motion       & Q3235565   & f/p      & 1           & hp    \\ \hline
radial velocity                    & Q240105    & f        & 1           & hp    \\ \hline
rest mass                          & Q96941619  & f/p      & 0            & hp    \\ \hline
speed                              & Q3711325   & p        & 2           & hp    \\ \hline
speed of light                     & Q2111      & f/p      & 1           & hp    \\ \hline
spherical pendulum                 & Q3299367   & p        & 1           & hp    \\ \hline
stress                             & Q206175    & f/p      & 1           & hp    \\ \hline
tangential acceleration            & Q2822927   & f/p      & 1           & hp    \\ \hline
tangential velocity                & Q103715245 & i/f/p    & 2           & hp    \\ \hline
uniform motion                     & Q376742    & f/p      & 1           & hp    \\ \hline
\end{tabular}}
\label{tab:SeedingList}
\end{table}

\subsection{Wikipedia Entity Linking}

Having seeded the missing Formula Concepts into Wikidata items, we can start linking them by transferring the annotations of our 25 selected Wikipedia articles\footnote{\url{https://github.com/ag-gipp/dataAnnoMathTex/tree/master/evaluation}}. In our first trial, we only add \verb|<math>| tag \verb|qid| attributes (see Section \ref{subsec:wikip.entitylink}) to equations (`=' sign in formula string). We skip second occurrences of the same QID in analogy to to the Wikipedia policy for natural language links (see the `Manual of Style'\footnote{\url{https://en.wikipedia.org/wiki/Wikipedia:Manual\_of\_Style}}). Our entity linking transfer script\footnote{\url{https://github.com/ag-gipp/AnnoMathTeX/tree/master/evaluation/wikipedia-export}}
employs Pywikibot to insert the qid links into the Wikitext of the respective articles by matching the formula \LaTeX strings. Before running it on the live articles, it was tested in a sandbox. At it's execution, the script spotted 508 candidates and skipping 263 duplicates (after first occurrence), finally 245 formulae were linked (with an OAuth token in the name of user `PhilMINT').
\subsection{Wikimedia Community Feedback}\label{subsec:wiki.comm.feedb}

To evaluate our data transfer pipeline, we attempt to answer the following research questions:
\begin{enumerate}
    \item What is the community acceptance of Wikidata item creation and population in terms of accepted or rejected changes and issue comments?
    \item What is the community acceptance of Wikipedia article formula entity links in terms of accepted or rejected changes and issue comments?
    \item Which issues are pointed out by the community? How can they be classified? 
\end{enumerate}

The following three subsections describe the reaction of the community to our seeding and linking experiment. We only discuss the issues, we considered most important. The full list can be found in the evaluation folder of the >>AnnoMathTeX<< repository.

\subsection{Community Feedback on Wikidata}\label{subsec:feedb.wikid}

As stated in Section \ref{subsec:form.conc.seed} and denoted in Table \ref{tab:SeedingList}, currently there are two common usages to seed identifier semantics (names and symbols): either using the Wikidata property `has part' (hp) or `calculated from' (cf).
However, at the moment, the Wikipedia special page for the formula semantics display only retrieves the information from `has part' properties and their symbols from `quantity symbol (string)' as opposed to `quantity symbol (LaTeX)', which was later created by the community.
In an effort to unite those usages to a standard that is compatible with the Wikipedia entity linking detail display, we started a discussion on the talk page of the opponent property `calculated from' (P934)\footnote{\url{https://www.wikidata.org/wiki/Property_talk:P4934}} with an appeal to use `has part' (P527) instead.
Concerns about the general validity of both properties were expressed. One community member responded that Wikidata should be usable independently of Wikipedia, not having to comply with the technical requirements for the special page display. It was pointed out that currently (as of December, 3rd 2020) in Wikidata of the items which have a formula, about 560 use `calculated from' (P4934) and about 150 use `has part' (P527).
Studying some sample equations with `calculated from' properties, another user pointed out cases, in which the usage of `calculated from does not seem reasonable.
Furthermore, the coexistence and different benefits of the properties `quantity symbol (string)' (P416), `quantity symbol (LaTeX)' (P7973), and `defining formula' (P2534) for the subexpression strings were discussed.

\subsection{Community Feedback on Wikipedia}\label{subsec:feedb.wikip}

Less than half a day after the execution of the script, two Wikipedia editors started a discussion on our user's talk page\footnote{\url{https://en.wikipedia.org/wiki/User_talk:PhilMINT}}. It was pointed out that the `defining formula' property of the corresponding Wikidata item is edited and evolving independently from the formula strings linked in the Wikipedia articles. Moreover, the Wikidata items need to be very specific to account for a particular formula, e.g., `kinetic energy of rotating body' (Q104145205). Sometimes a disambiguation is needed to distinguish the mathematical terms from other word meanings, e.g., `work' (Q42213) meaning energy transfer vs. `work' (Q6958747) meaning labour. In some articles, a one-line formula includes several sub formulae with different meanings that would need three different QIDs. Lastly, it was proposed that the special pages and corresponding Wikidata items should have a `what links here' display. This way dependencies could be analyzed, and editors warned.

\subsection{Curating a Goldstandard}\label{subsec:cur.gold}

Following a suggestion of an experienced Wikipedia user, we created an annotation table to discuss our contributions at our Talk page\footnote{\url{https://en.wikipedia.org/wiki/User\_talk:PhilMINT\#QID\_annotation\_table}}.
To persist our contributions as a Goldstandard, we inserted our seeding list (Table \ref{tab:SeedingList}) into the benchmark MathMLben~\cite{DBLP:conf/jcdl/SchubotzGSMCG18}, which provides an open-access user interface\footnote{\url{https://mathmlben.wmflabs.org}}. The seeding list inserts range from Gold ID 310 to 375.

\section{Conclusion \& Outlook}\label{sec:concl.outl}

In this section, we summarize our contributions and outline the benefits, challenges, and future directions of our work.

\subsection{Conclusion}\label{subsec:concl}

In this paper, we evaluated the document annotation speedup for and Wikimedia community acceptance of Mathematical Entity Linking. We selected 25 articles from physics (classical mechanics of motion), containing a total of 1797 formulae. We identified `Wikipedia' and the `word window' as best sources for identifier and formula name recommendations, respectively. Using the >>AnnoMathTeX<< AI assistance, we were able to speed up the annotations by a factor of 1.4 for formulae and 2.4 for identifiers, respectively. We transferred 245 formula linkings to the Wikipedia articles and contributed to 42 Wikidata items to ground the formula semantics. The community rejected 12\% of the edited Wikipedia articles and 33\% of the Wikidata items within the first month. We persisted the Formula Concepts seeding list (Wikidata items that were annotated and linked) into the benchmark MathMLben~\cite{DBLP:conf/jcdl/SchubotzGSMCG18}.

\subsection{Outlook}\label{subsec:outl}

\paragraph{Benefits}

Performing entity linking in Wikipedia articles, we now have collected different representations for a number of Formula Concepts, which can be used as training data for Formula Concept Recognition (FCR).
Creating labeled datasets of annotated documents and Formula Concept benchmarks will be an important step for mathematical language processing towards making mathematical content machine-interpretable. This means that IR systems, such as semantic search and question answering as well as ML algorithms (e.g., for document classification or clustering), can exploit the data.
Furthermore, a `Formula Graph Database' can be constructed to visualize relationships or dependencies between formulae and identifiers and automated reasoning. In analogy to Google's `PageRank' for popularity ranking of webpages, a `FormulaRank' for popularity assessment of formulae will be feasible.

\paragraph{Challenges}

As pointed out by the community, a major drawback of the Wikipedia special pages is that the data they display, i.e., the Wikidata items, evolve independently from the articles. To address this, the data should be persisted, such that the formula with its annotations in the item still matches the one that is linked in Wikipedia.
Furthermore, because disambiguation of formula and identifier items (e.g., deciding between work = energy transfer and work = labour) needs human supervision, the seeding can not be fully automated. Thus, it would be beneficial to directly integrate the annotation recommender system into the visual editor of Wikipedia with seeding connection to Wikidata.
Our experiments showed that the Wikimedia integration of the seeding and entity linking is very important, since the time effort to seed was almost twice as large than to annotate.
Lastly, we could not evaluate whether accepting recommendations deteriorate the quality of the annotations compared to manual inserts. Maybe a lazy annotator is inclined to click rather than type.

\paragraph{Future Work}

In the future, we will extend our elaborations of an annotation standard and guidelines and discuss it with the community. 
Apart from the community feedback on entity links in Wikipedia and item seeding in Wikidata, we plan to carry out a user survey for >>AnnoMathTeX<< in preparation for the Mediawiki integration.
For the integration, we will build a >>MathWikiLink<< API to provide the annotation recommendation sources and add the constraint that it is only possible to add a link between Wikipedia and Wikidata if the defining formula matches.
Moreover, there will be a feature that displays the special page formula information on Wikipedia in a popup\footnote{\url{https://phabricator.wikimedia.org/T208758}} as currently only available for Wikilinks and references.
For the manual typing insertions, an auto-completion will be built (considering a combination of recommendations from all sources). The system needs to learn to improve with increasing user interactions and annotation contributions from the community. Therefore, a reinforcement ranking, pushing frequently accepted recommendations higher, will be employed.
Finally, the mathematical entity links can also be used to build a formula reference system for `math citations'.

\section*{Acknowledgment}

This work was supported by the German Research Foundation (DFG grant GI-1259-1).

\printbibliography[keyword=primary]
\end{document}